\documentclass[12pt]{iopart}

\usepackage{graphicx}
\usepackage{dcolumn}
\usepackage{bm}

\usepackage{amssymb}
\usepackage{color}

\def\be{\begin{eqnarray}}
\def\ee{\end{eqnarray}}
\def\ba{\begin{array}}
\def\ea{\end{array}}

\begin{document}


\title[Current-induced asymmetries in narrow Hall bars]{Current-induced
  asymmetries of incompressible stripes in narrow quantum Hall systems}

\author{Rolf R. Gerhardts, Konstantinos Panos, and J{\"u}rgen Weis}
\address{Max-Planck-Institut f\"ur Festk\"orperforschung, 
Heisenbergstrasse 1, D-70569 Stuttgart, Germany}%

\ead{R.Gerhardts@fkf.mpg.de}


\begin{abstract}
We present recent experimental results confirming previously predicted strong
asymmetries of 
the current distribution in narrow Hall bars under the conditions of the
integer quantum Hall effect (IQHE). Using a previously developed
self-consistent screening and transport theory of the IQHE, we
investigate how these asymmetries, which are due to a non-linear feedback
effect of the imposed current on the electron distribution in the sample,
 depend on relevant parameters, such as
the strength of the imposed current, the magnetic field, the
 temperature, and the collision broadening of the Landau-quantized energy
 bands. We find that many aspects of the experimental results can be understood
 within this approach, whereas other aspects require explicit consideration of
 mechanism, which enforce the breakdown of the IQHE.
\end{abstract}

\submitto{\NJP}
\maketitle

\section{\label{sec:1} Introduction}
Scanning force microscopy (SFM)  \cite{Weitz00:349} has provided interesting
information \cite{Weitz00:247,Ahlswede01:562,Ahlswede02:165}
about the position dependence of Hall potential and current
density in narrow Hall bars, with a width of about $10\,\mu$m. Under
strong perpendicular magnetic fields $B$, which allow the observation of the
integer quantized Hall effect (QHE) \cite{vKlitzing80:494}, three types of the
spatial variation of the Hall potential were observed. For $B$
values well outside the plateaus of the QHE, the Hall potential varied
essentially linearly across the sample (``type I''), as one expects from the
classical Drude theory of the Hall effect.
As the $B$ field entered a plateau region from the high-field side,   a
non-linear potential drop in a broad region in the middle of the sample was
observed (``type II''). Most interestingly, near the low-$B$ edges of the 
QH plateaus the Hall potential was constant in a broad region in the center of
the sample  and dropped across two strips that moved with decreasing $B$
 towards the sample edges (``type
III'') \cite{Ahlswede01:562}.  Position and width of these strips coincided 
 with those of the incompressible strips (ISs) expected to form in the sample
 due to the non-linear screening properties of the two-dimensional electron
 system (2DES)  in strong magnetic fields
\cite{Chklovskii92:4026,Chklovskii93:12605,Lier94:7757,Oh97:13519}.
In the ``perfect screening'' limit \cite{Chklovskii92:4026,Chklovskii93:12605}
 of vanishing temperature and collision broadening, the potential varies only in
 these ISs, where the local filling factor is an integer and the electron
 density is constant, and in between the ISs the potential is flat.

These experimental results have been qualitatively reproduced by a
self-consistent calculation \cite{Guven03:115327} of the screened 
confinement potential $V({\bf r})$ and the electron and current densities
under the assumption, that the imposed, and in general dissipative, current
density  $\mathbf{j}({\bf r})$ obeys a local form of Ohm's law,
$\mathbf{j}({\bf r})= \hat{\sigma}({\bf r}) \mathbf{E}({\bf r})$, 
 with the gradient of the electrochemical potential
$\mu^{\star}(\bf r)$ as driving electric field, ${\bf E}=\nabla \mu^{\star}/e$.
 The position-dependent conductivity tensor  $\hat{\sigma}({\bf r})$,
 determining longitudinal and Hall current,      was,
in turn, calculated from  $V({\bf r})$ and $\mu^{\star}(\bf r)$ under the
assumption of {\em local thermodynamic equilibrium} \cite{Guven03:115327}. 
Interestingly, these self-consistent calculations
showed an unexpected non-linear feedback effect of the imposed current on 
the electron density: the calculated ISs near the sample edges, and as a
consequence the current distribution among and the drop of the Hall potential
across these strips, became with increasing strength of the
imposed current increasingly asymmetric. However, since this
asymmetry, which vanishes in the linear-response limit of low imposed
current \cite{Siddiki04:195335},  was not clearly seen in the experiments,
it was not investigated in detail.

\begin{figure}[h]
\noindent \begin{minipage}[b]{0.58\linewidth}
  \includegraphics[width=1.25\linewidth,angle=0]{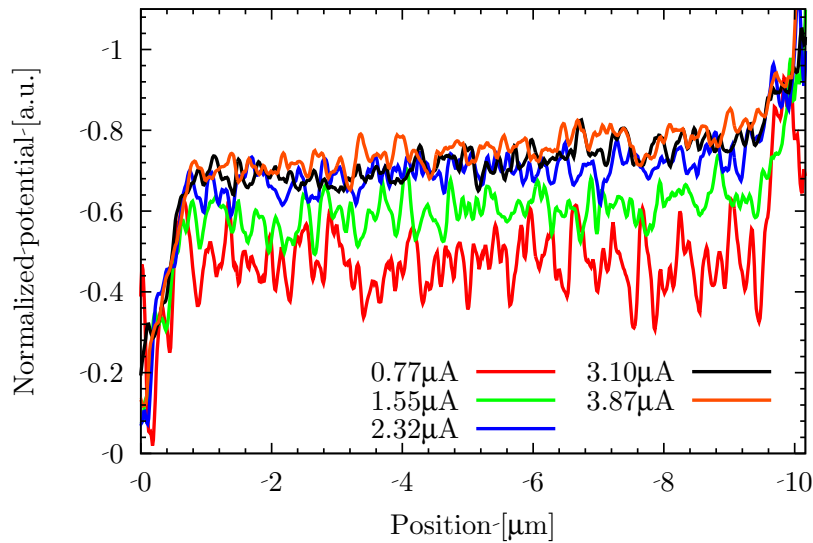}
\end{minipage} \hfill 
 \begin{minipage}[b]{0.42\linewidth}
\caption{\small \label{fig:edgedom} Normalized Hall potential
  measured 
 on a $10\,\mu$m wide Hall bar with electron density $n_{\rm
    el}=2.9\cdot 10^{15}$m$^{-2}$ at $B=5.6\,$T (i.e., 
  average Landau   level filling factor $\nu=2.14$). The asymmetry increases
  with increasing strength of the imposed current.  From 
  Ref.~\cite{Panos11:thesis}.}
\end{minipage}
\end{figure}

In a recent series of similar experiments, SFM was  applied to
investigate the breakdown of the QHE in narrow Hall bars under strong imposed
currents \cite{Panos11:thesis}. Unidirectional current pulses were used,
to avoid averaging out of effects depending on the direction of the imposed
current. Typical results, corresponding to the edge-determined ``type
III'' behavior of 
Ref.~\cite{Ahlswede01:562}, are shown in Fig.~\ref{fig:edgedom}.
Here the asymmetry of the steps in the Hall potential, caused by the
current-carrying strips near the sample edges, is clearly seen, and the
asymmetry increases with increasing strength of the imposed current.
Experimental details and results for the ``bulk-determined'' situation
(corresponding to the ``type I'' and ``type II'' behavior of
Ref.~\cite{Ahlswede01:562}) will be published
separately \cite{Panos13:unpub}. 
In view of these new experiments it seems interesting to investigate in some
detail the origin of these asymmetries, and their dependence on
relevant parameters like temperature, strength of the imposed current, 
collision-broadening of the Landau levels, and, of course, the magnetic field
value, which is important for existence and position of the ISs.

We should mention that current-induced asymmetries have also been predicted
\cite{Siddiki09:17008} for, and experimentally observed \cite{Siddiki10:113011}
in asymmetric Hall bars, with different profiles of
confinement potential, and therefore electron density, at opposite edges.
 In such samples the shape of the quantized Hall
plateaus changes, if the direction of the imposed current is reversed. These
experiments indicate current-induced asymmetries, but provide no detailed
information about the local distribution of the imposed current.

\section{\label{sec:2} Model}
In view of the unavoidable intrinsic fluctuations of the experimental
data (see Fig.~\ref{fig:edgedom} and Ref.~\cite{Ahlswede01:562}), we
will not attempt to reproduce them quantitatively, but only to achieve a
qualitative understanding, and we will try to keep the model simple and the
computer times short, following
closely model assumptions and methods used in previous
work \cite{Lier94:7757,Oh97:13519,Guven03:115327,Siddiki04:195335}. 
We have in mind a GaAs/(AlGa)As heterostructure containing a quasi 2DES,
populated by a doping layer and laterally confined by top gates, but we
assume the distances between the layers to be much smaller than the width of
the 2DES, and neglect them. Thus we assume that all charges, including
induced gate-charges, reside in the plane
$z=0$ \cite{Chklovskii92:4026,Chklovskii93:12605,Lier94:7757,Oh97:13519},   
and we model the Hall bar as a 2DES in that plane, being translation
invariant in $y$ direction and confined in $x$ direction to  the
stripe $|x|<d$. We neglect exchange and correlation effects and
spin-splitting, and we replace the mutual 
Coulomb interaction between the electrons by an effective Hartree potential,
determined by the density  $n_{\rm el}(x)$ of the 2DES. With the
boundary conditions $V(x,z)\rightarrow 0$ for $|z|\rightarrow \infty$ and
$V(x,0)=V_L$ for $x<-d$ and $V(x,0)=V_R$ for $x>d$ the potential $V(x,z)$
is obtained as an integral over the charge density in  the stripe
$|x|<d$ \cite{Guven03:115327}. A constant density $n_D$ of ionized donors
(Si$^+$) in $|x|<d$ contributes $V_D(x)=-E_D[1-(x/d)^2]^{1/2}$, with $E_D=2 \pi
e^2n_Dd/\bar{\kappa} $, to  the effective confinement potential
$V(x,0)=V(x)=V_D+V_g+V_H$, a gate voltage $V_R-V_L$ contributes
$V_g(x)=(V_L+V_R)/2+(V_R-V_L)\,\arcsin(x/d)/\pi$, and
$V_H(x)=(2e^2/\bar{\kappa})\int_{-d}^d dx'\, K(x,x')  
n_{\rm   el}(x')$ is the Hartree potential,   determined by the kernel
 \be K(x,x')=\ln
\left|\frac{\sqrt{(d^2-x^2)(d^2-x'^2)}+d^2-x'x}{(x-x')d} \right| \,.
  \label{eq:kernel-inplane}  \ee
Here $V(x)$ is normalized to give the potential energy of an electron, and 
$\bar{\kappa} $ is an effective dielectric
constant \cite{Guven03:115327,Siddiki04:195335}.  

To obtain self-consistency, we have to calculate  $n_{\rm el}(x)$ from
$V(x)$. The full Hartree approximation requires to solve the eigenvalue
problem with the potential $V(x)$, but we assume that $V(x)$ varies within the
2DES slowly on
the scale of the magnetic length $\ell_B=\sqrt{\hbar c/eB}$. This allows to
approximate the energy eigenvalue of the Landau state with center  coordinate
$X$ by $E_n(X)=V(X)+E_n$, with  $E_n$ the Landau energy
without confinement, and to neglect the extent of the states in $x$-direction,
when calculating the density \cite{Gerhardts08:378}. This leads in thermal
equilibrium to the Thomas-Fermi approximation 
\be \label{eq:TFA}
n_{\rm el}(x)=\int dE \, D(E)\, f([E+V(x)-\mu^{\star}]/k_BT),
\ee
with $f(\varepsilon)=1/[1+\exp(\varepsilon)]$ the Fermi function, $k_B$
Boltzmann's constant, $T$ the temperature, and $D(E)$ the density of states
(DOS) in
the absence of the confinement potential. We take $D(E)=D_0=m/(\pi \hbar^2)$
for $B=0$, and for large, quantizing magnetic fields
$D(E)=\sum_{n=0}^{\infty}A_n(E)/(\pi \ell_B^2)$ with the Gaussian form
$A_n(E)= \exp(-[E_n-E]^2/\Gamma^2)/(\sqrt{\pi} \, \Gamma)$ of the spectral
function \cite{Gerhardts75:285}, with  $\Gamma=\gamma \, \hbar \omega_c
[10$T$/B]^{1/2}$ and $\gamma$ a measure of the collision broadening.
Here $E_n=\hbar \omega_c(n+1/2)$ are the Landau energies with
$\omega_c=eB/mc$ the cyclotron frequency, $m$ is the effective mass, and we
assume spin-degeneracy. 
\begin{figure}[t]
\noindent \begin{minipage}[b]{0.58\linewidth}
  \includegraphics[width=1.1\linewidth,angle=0]{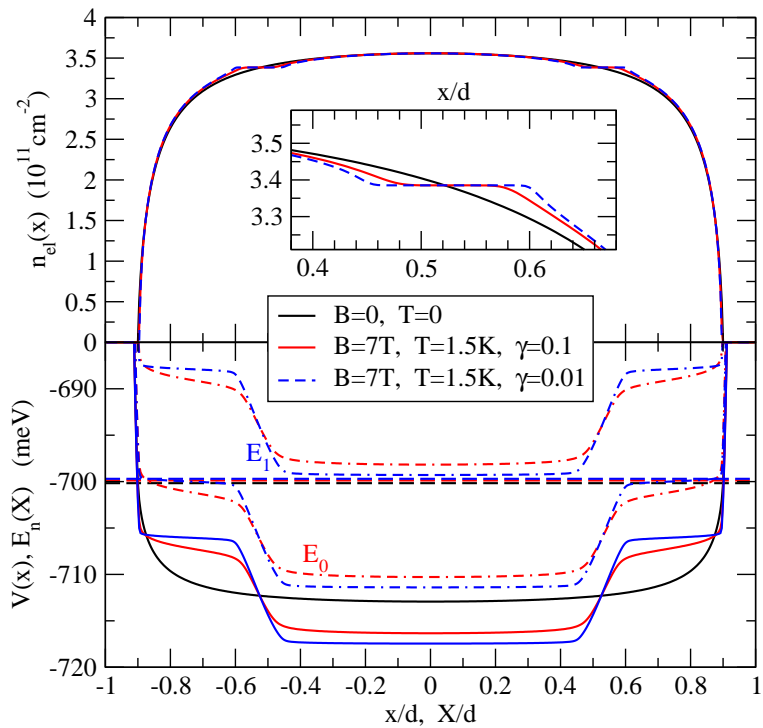}
\end{minipage} \hfill
\begin{minipage}[b]{0.42\linewidth}
\caption{ \label{fig:v0dens} \small Self-consistent potential and
  electron density in thermal equilibrium. Black lines: $B=0$, $T=0$; other
  lines for   $B=7\,$T, $T=1.5\,$K, and Gaussian spectral functions of width
 $\Gamma=\gamma \, \hbar \omega_c [10$T$/B]^{1/2}$, with
$\gamma=0.1$ (red), and   $\gamma=0.01$ (blue), respectively;  broken lines:
$\mu^{\star}$,   dash-dotted: Landau energies.}
\end{minipage}
\end{figure}
Figure~\ref{fig:v0dens} shows typical thermal-equilibrium results for 
 a sample  with gate distance $2d=3\mu$m
and with $N\sim 500$ equidistant mesh points for the evaluation of integrals.
(Results for $d=2.5\,\mu$m and $N\sim 1000$ are very similar). In view of the
experiments \cite{Panos11:thesis}, we have chosen  a donor
density of $n_D=4\cdot 10^{11}$cm$^{-2}$, so that with the GaAs values
$m=0.067\, m_0$ and $\bar{\kappa}=12.4$ the minimum of $V_D(x)$ is
$V_D(0)=-E_D=-4.38$eV. For $B=0$ and $T=0$ we confine the electron
density to the stripe $|x|<0.9\,d$, solve the resulting linear
integral equation \cite{Guven03:115327},  and obtain a symmetric
density profile with an average electron density
$\bar{n}_{\rm el}=2.90\cdot 10^{11}$cm$^{-2}$.  The resulting confinement
potential vanishes on the gates,
$V_L=V_R=V(\pm d)=0$ and is well screened within the 2DES, with minimum
$V(0)=-0.71\,$eV, and the electrochemical potential is $\mu^*=-0.70\,$eV,
so that the total potential varies within the 2DES only by about $10\,$meV
 (see black lines of Fig.~\ref{fig:v0dens}).
 Starting with these values, the Newton-Raphson method with stepwise
 temperature changes \cite{Guven03:115327} was used to obtain results for high
 magnetic 
 fields and low temperatures.  The pinning of the Landau levels to
 the constant electrochemical potential  becomes more effective for smaller
 $\gamma$  and, of course, for lower $T$. The current $-e \langle
 n,X|\hat{v}_y|n,X\rangle =(e/m\omega_c)dE_n(X)/dX $ carried by the state
 $|n,X \rangle$ reduces in our approximation to $(c/B)dV(X)/dX$, so
 that the intrinsic (Hall) currents in the perfect screening limit are
 confined to the incompressible strips, with opposite current directions near
 opposite sample edges, and vanish in the compressible regions
 of constant potential. 

To describe a {\em stationary non-equilibrium state} with an imposed current $I$
along the Hall bar, we assume {\em local equilibrium} and replace the
constant electrochemical potential $ \mu^{\star}$ 
in Eq.~(\ref{eq:TFA}) by a position-dependent one, $ \mu^{\star}(x,y)$.
As in Ref.~\cite{Guven03:115327} we
assume, that the current density obeys Ohm's law with the driving
electric field ${\bf  E}\equiv  {\bf \nabla} \mu^* /e$ and 
position-dependent conductivity tensor components
$\sigma_{xx}(x)=\sigma_{yy}(x)=\sigma_{\ell}(x)$ and
$\sigma_{yx}(x)=-\sigma_{xy}(x) =\sigma_H(x)$. To maintain translation
invariance in $y$-direction,   ${\bf  E}(x,y)$ must be independent of $y$. In
a stationary state we have  
 $ {\bf \nabla} \cdot{\bf j} =0$ and ${\bf \nabla} \times {\bf E}={\bf 0}$, so
 that the current density $j_x$ across  and the field $E_y$ along the Hall bar
 are independent of $x$, $j_x(x) \equiv j_x^0$ and $E_y(x)\equiv E_y^0 $. Then
 the electrochemical potential assumes the form $ \mu^{\star}(x,y)=
 \mu^{\star}(x)+e E_y^0 y$ and, in order to guarantee the translation
 invariance of the electron density, we must add the term $e E_y^0 y$ also to
 the confinement potential (where it describes the effect of a source-drain
 voltage along the Hall bar), so that the difference $V(x,y)- \mu^{\star}(x,y)
 =V(x)-\mu^{\star}(x)$ is independent of $y$.

In the following we assume sufficiently wide depletion strips between the
2DES and the gates, so that the current across the Hall bar vanishes, $
j_x^0=0$. Then the remaining current and field components are easily expressed
in terms of the resistivity components
$\rho_{\ell}(x)=\sigma_{\ell}(x)/[\sigma_{\ell}(x)^2+\sigma_H(x)^2]$  and 
$\rho_H(x)=\sigma_H(x)/[\sigma_{\ell}(x)^2+\sigma_H(x)^2]$ as
\be  \label{jE2}
j_y(x)=E_y^0/\rho_{\ell}(x), \qquad E_x(x)=\rho_H(x)\, j_y(x),
\ee
resulting in $\mu^*(x,y)=e E_y^0 y+ \mu^*(x)$ with
\be \label{mudiff}
  \mu^*(x)= \mu^*_0 +e E_y^0 \int_0^x
dx' \, \rho_H(x')/\rho_{\ell}(x')  \,. \ee 
Here  $\mu^*_0$ occurs as constant determining the average electron density,
and $E_y^0$ is determined by the relation $\int_{-d}^d dx j_y(x)=I$ as
$E_y^0=I/ \int_{-d}^d dx \,[1/\rho_{\ell}(x)]$. If in the limit $T\rightarrow
0$ an IS develops where $\rho_{\ell}(x)\rightarrow 0 $, the last integral
diverges, i.e., the electric field and the resistance along the Hall bar
vanish, and the sample will show the QHE.

For the sake of comparison, we recall the classical Drude model
\be \label{Drude}
\rho_{\ell}^D(x)=m/[e^2\tau n_{\rm el}(x)], \quad \rho_H^D(x)=\omega_c \tau
\, \rho_{\ell}(x), \ee
with a constant scattering time $\tau$, which according to Eq.~(\ref{jE2})
leads to a current density proportional to the electron density, $j_y^D(x)
\propto n_{\rm el}(x)$, and to a constant Hall field $E_x^D=\omega_c \tau E_y^0$.

In the following we include Landau quantization and Shubnikov-de~Haas effect
by taking as model for the conductivity tensor \cite{Guven03:115327}
\be \label{sigma}
\sigma_H(x)&=&(e^2/h) \nu(x), \qquad \nu(x)=2\pi \ell_B^2 n_{\rm el}(x),\\
\sigma_{\ell}(x)\!&=&\!\frac{e^2}{h}\! \int_{-\infty}^{\infty}\!\!dE\,[-f'_E]
\sum_{n=0}^{\infty}
(2n+1)[\sqrt{\pi}\Gamma A_n(E)]^2, \nonumber
\ee
where $ n_{\rm el}(x)$ is calculated according to Eq.~(\ref{eq:TFA}) with the
argument of the Fermi function replaced by $[E+V(x)-\mu^*(x)]/k_BT$ and where
$f'_E=\partial f([E+V(x)-\mu^*(x)]/k_BT)/\partial E$. To obtain reasonable
results for the Hall voltage, we follow Ref.~\cite{Guven03:115327} and
require that the induced charges on each gate
remain constant, independent of the imposed current $I$. For the gate at $x>d$
this requires to fix \cite{Guven03:115327}
\be \label{QR}
Q_R=\frac{2e}{\pi}\!\int_{-d}^d\!\! dx\,  [n_D\!-\!n_{\rm el}(x)]
\arctan \sqrt{\frac{d\!+\!x}{d\!-\!x}}. 
\ee

We modify the previous approach \cite{Guven03:115327} in two respects.
First, we avoid the cutoff   $\sigma_{\ell}(x)>\epsilon \cdot \sigma_H(x)$
with   $\epsilon =10^{-4}$ used in  Ref.~\cite{Guven03:115327}, since it
becomes crucial already for relevant values of temperature and collision
broadening, and makes the results for the current density in the incompressible
strips unreliable.
The second modification has been introduced \cite{Siddiki04:195335}
and discussed  \cite{Gerhardts08:378,Siddiki09:17008} previously. The
Thomas-Fermi approximation (TFA), 
which neglects the spatial extent of the Landau eigenstates, 
and the assumption of a strictly local form of Ohm's law, which is reasonable
on a length scale much larger than the average electron separation, lead to
physical and mathematical problems.  A simple and reasonable way to overcome
these problems is to replace in Ohm's law the conductivity tensor
$\hat{\sigma} (x)$, calculated
within the local thermal equilibrium approach, by a smoothed form
\cite{Siddiki04:195335,Gerhardts08:378}  
\be \label{aversigma} 
\hat{\bar{\sigma}}(x)=\frac{1}{2\lambda} \int_{-\lambda}^{\lambda} d\xi \,
\hat{\sigma}(x+\xi).
\ee
We will investigate the effect of this averaging in addition to that of
temperature, collision broadening, current strength, and magnetic field.

\section{\label{sec:3} Results}

Low-temperature results with imposed current were obtained following two
routes. In the first, we start with the equilibrium state ($I=0$) at low
temperature and increase the current stepwise. For given $I$, we first keep
$\mu^*(x)-\mu^*_0$ fixed and calculate  $V(x)$ and $n_{\rm el}(x)$ as
in equilibrium, determining $\mu^*_0$ and $V_R-V_L$ so that the average
electron density $\bar{n}_{\rm el}$ and the induced gate charge $Q_R$,
Eq.~(\ref{QR}), remain unchanged. Then we calculate for the resulting
$V(x)-\mu^*(x)$  the conductivities, Eqs.~(\ref{sigma}), and via Ohm's law and
Eq.~(\ref{mudiff}) a new $\mu^*(x)-\mu^*_0$. This procedure is iterated until 
the changes of $V(x)$ and $\mu^*(x)$ are negligible. The next value of $I$ 
is treated in the same way. 

\begin{figure}[h]
\noindent \begin{minipage}[b]{0.58\linewidth}
  \includegraphics[width=1.2\linewidth,angle=0]{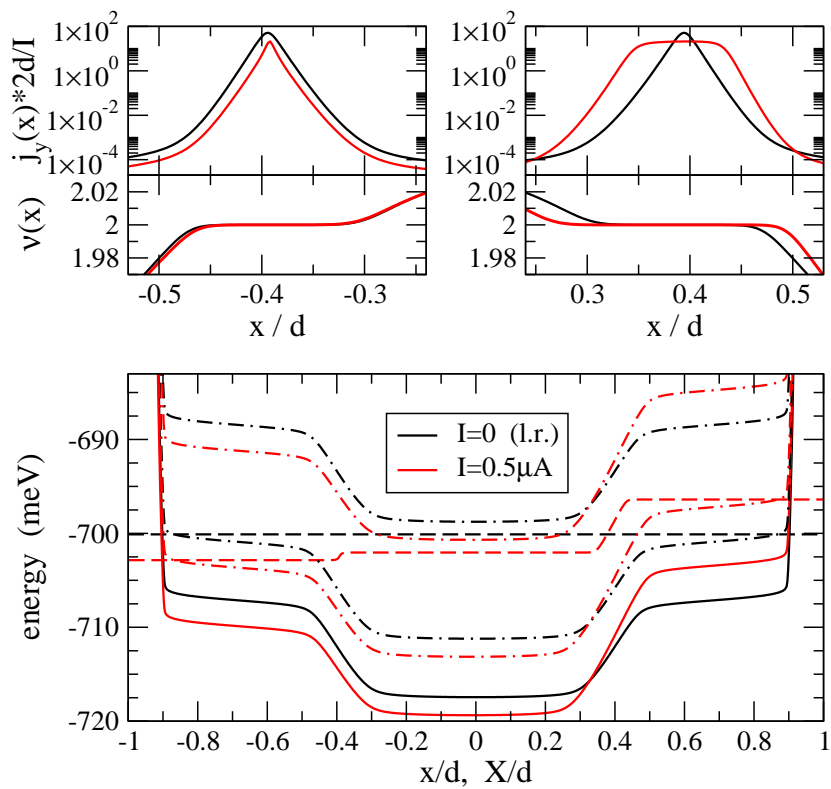}
\end{minipage} \hfill
\begin{minipage}[b]{0.42\linewidth}
\caption{ \label{fig:vnujy} \small Lower part: self-consistent
  potential $V(x)$ (solid lines),   electrochemical potential
  $\mu^{\star}(x)$ (broken lines), and Landau
  energies $E_n(X)$ (dash-dotted); upper parts:  density $j_y(x)$ of imposed
  current and 
  filling factor   $\nu(x)$ near the incompressible strips; $B=7.2\,$T,
  $T=4\,$K,   $\gamma=0.01$, $N=980$.}
\end{minipage}
\end{figure}

\begin{figure}[h]
\noindent \begin{minipage}[b]{0.58\linewidth}
  \includegraphics[width=1.15\linewidth,angle=0]{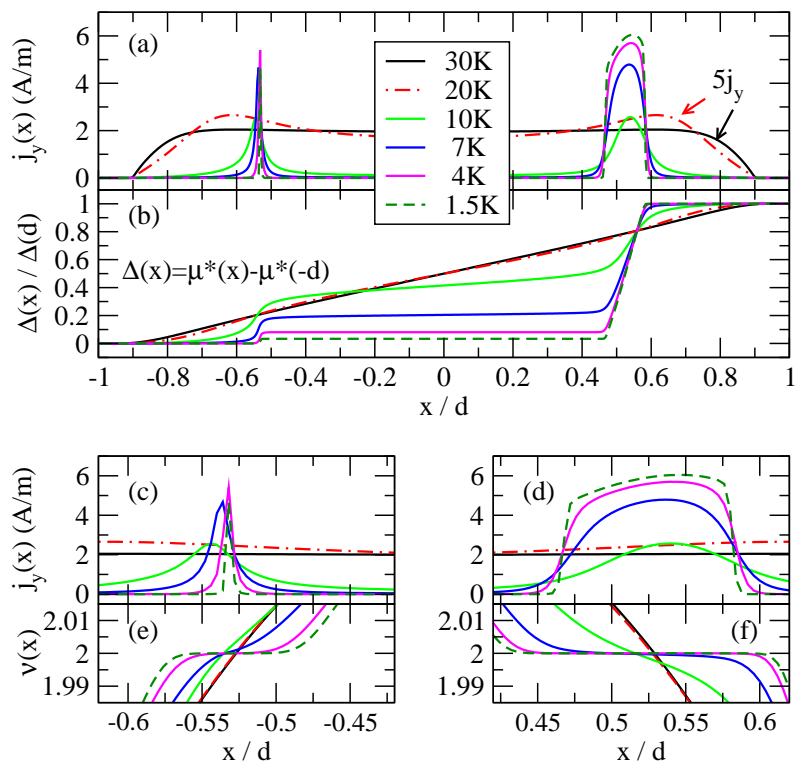}
\end{minipage} \hfill
\begin{minipage}[b]{0.42\linewidth}
\caption{ \label{fig:TdepB7i1} \small Position dependence of 
  current density $j_y(x)$, (a), and normalized Hall potential, (b), for
  different temperatures, from $T=30$K to $T=1.5$K. Details of
  $j_y(x)$ and of $\nu(x)$ near the positions of
  incompressible strips are shown in (c) and (d), and in (e) and (f),
  respectively; $B=7\,$T, $\Gamma/\hbar \omega_c=0.12$, imposed current
  $I=1\,\mu$A. }
\end{minipage}
\end{figure}

\begin{figure}[h]
\noindent \begin{minipage}[b]{0.6\linewidth}
  \includegraphics[width=1.1\linewidth,angle=0]{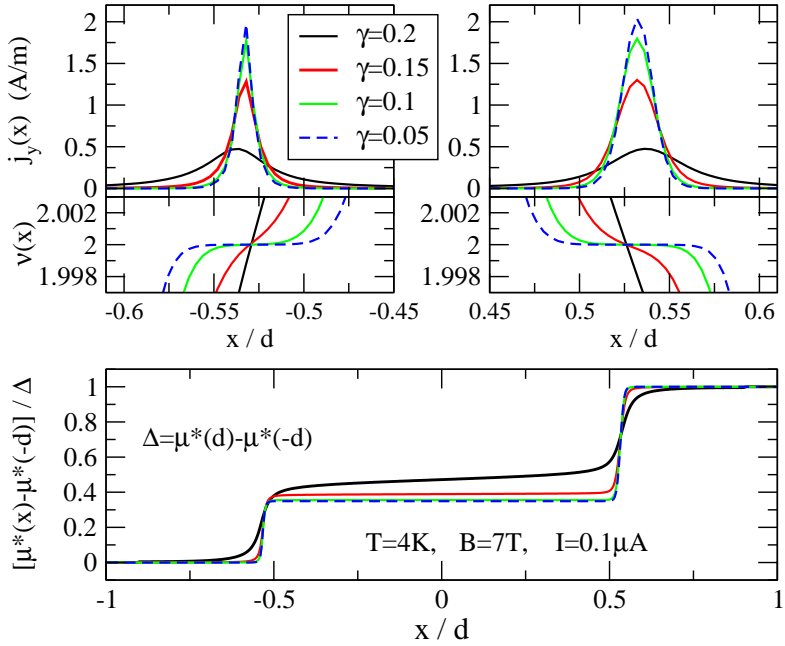}
\end{minipage} \hfill
 \begin{minipage}[b]{0.4\linewidth}
\caption{ \label{fig:gamdep} \small Normalized Hall
  potential, current density $j_y(x)$, and  $\nu(x)$ as in
  Fig.~\ref{fig:TdepB7i1}, 
  but for $T=4\,$K, $I=0.1\,\mu$A, and different values of collision
  broadening.}
\end{minipage}\end{figure}

\begin{figure}[h]
\noindent \begin{minipage}[b]{0.6\linewidth}
  \includegraphics[width=1.1\linewidth,angle=0]{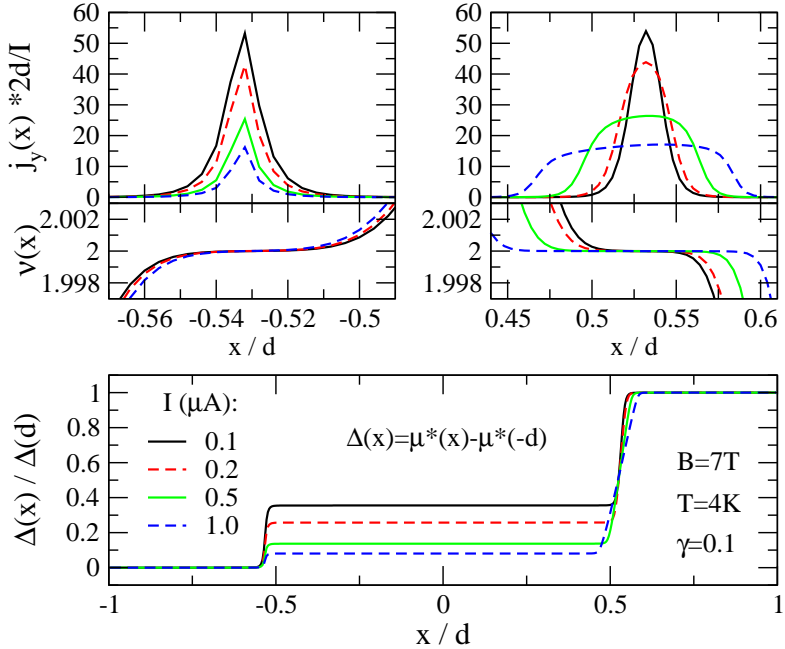}
\end{minipage} \hfill
 \begin{minipage}[b]{0.4\linewidth}
\caption{ \label{fig:Idep}\small Normalized Hall
  potential,  $j_y(x)$, and  $\nu(x)$ as in Fig.~\ref{fig:TdepB7i1},
  but for $T=4\,$K, $\gamma=0.1$, and different values of the imposed current
  $I$.}
\end{minipage}
\end{figure}

Figure~\ref{fig:vnujy} shows corresponding results
for $I=0$ and $I=0.5\,\mu$A at $T=4\,$K. Similar results have been presented
in Ref.~\cite{Siddiki09:17008}, but without details of the
current density  $j_y(x)$, which decreases by several orders of magnitude
already inside the ISs.
Note the logarithmic scale for  $j_y(x)$, which shows the
symmetric linear response result for $I=0$ and the strongly asymmetric result
for $I=0.5\,\mu$A. Outside the ISs $j_y(x)$ is at least  about
five orders of magnitude smaller than in their centers. As a consequence, the
``Hall potential'' $\mu^{\star}(x)$ is practically constant outside the ISs.
In the IS near
$x=0.4\,d$ the imposed current has the same direction as the intrinsic 
current. This leads to an increase of the potential variation across 
 and a broadening of this strip, whereas the IS near $x=-0.4\,d$
is much less affected. As a consequence, the variation of the Hall potential
across the IS near $x=0.4\,d$ is larger than that across the IS near
$x=-0.4\,d$. Since we prohibit charge transfer between the gates, i.e. keep
$Q_R$ constant, a gate voltage $(V_R-V_L)/e= [V(d)-V(-d)]/e= 10.795\,$mV builds
up, which is considerably larger than the calculated Hall voltage
$U_H=[\mu^{\star}(d)- \mu^{\star}(-d)]/e=6.4532\,$mV,  satisfying
$U_H=R_H(2) I$ with the quantized Hall resistance $R_H(2)\!=\!
0.5\,h/e^2\, \widehat{=} 
\, 12.9064\,$k$\Omega$ at filling factor $\nu=2$. These results depend, of
course, on the direction of the imposed current $I$. 
Since the considered sample is symmetric without imposed
current, $n_{\rm el}(-x)= n_{\rm el}(x)$, $V(-x)=V(x)$, etc., we obtain
for the symmetry due to the non-linear feedback of $I$
\be \label{sym-I}
n_{\rm el}(x;-I)=n_{\rm el}(-x;I), \quad j_y(x;-I)=-j_y(-x;I).
\ee
The normalized Hall potential $H(x)=\Delta(x)/\Delta(d)$ with $
\Delta (x)=\mu^{\star}(x)-\mu^{\star}(-d) $, plotted in
Figs.~\ref{fig:TdepB7i1} - \ref{fig:Idep}, 
changes under inversion of $I$ as $H(x;-I)=1-H(-x;I)$. 

Another route to the same results is to
 introduce the current at high temperature and then to lower $T$ stepwise,
 keeping $B$ and $I$ constant.
A typical result is shown in Fig.~\ref{fig:TdepB7i1} for a relatively high
imposed current, $I=1\,\mu$A. At sufficiently
high temperature, $T\gtrsim 30\,$K, the current density $j_y(x)$ is
proportional to the electron density $n_{\rm el}(x)$, which drops
monotonously from the center towards the edges, and the Hall voltage,
Fig.~\ref{fig:TdepB7i1}(b), varies linearly across the sample.
At low temperature, $T\lesssim 4\,$K, $n_{\rm el}(x)$ develops very
asymmetric plateaus with local filling factor $\nu(x)=2$, see
Fig.\ref{fig:TdepB7i1} (e) and (f), and   $j_y(x)$ is confined to
 these plateaus, where $\sigma_{\ell}(x)$ and
therefore the dissipation becomes small. As a consequence, the Hall
potential drops asymmetrically  across these plateaus and becomes
constant outside.


Figure~\ref{fig:gamdep} shows, for a weaker current,
that the asymmetry decreases with increasing collision broadening.
Apparently the current is
already largely concentrated in the regions with local filling factor
$\nu(x) \approx 2$ before pronounced plateaus and ISs develop. As
shown in Fig.~\ref{fig:Idep}, the width of the IS and the $j_y(x)$
peak near $x/d=0.53$ increase considerably with increasing imposed
current $I$, whereas the corresponding width near  $x/d=-0.53$ shrinks
only a little (note the different scales of the left and the right
parts of Fig.~\ref{fig:Idep}).

\begin{figure}[t]
\noindent \begin{minipage}[b]{0.59\linewidth}
  \includegraphics[width=1.2\linewidth,angle=0]{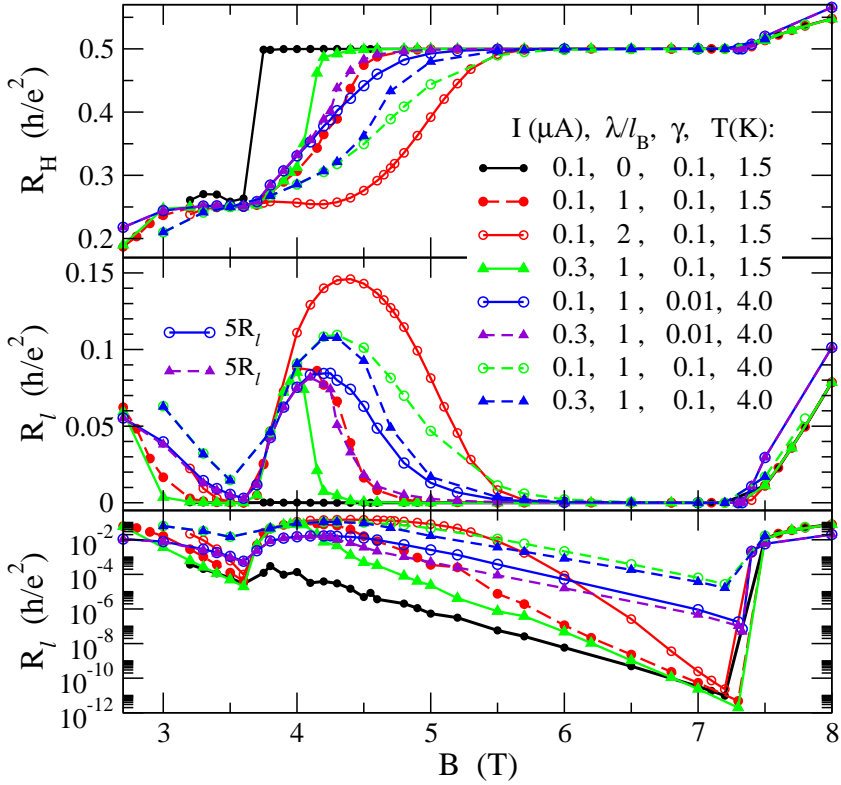}
\end{minipage} \hfill
\begin{minipage}[b]{0.4\linewidth}
\caption{ \label{fig:RlHvsB} \small $B$-dependence of Hall resistance
 $R_H=[\mu^{\star}(d)- \mu^{\star}(-d)]/(eI)$  (upper part) and longitudinal
 resistance $R_{\ell}=E_y^02d/I$ in linear (medium) and logarithmic scale
 (lower part), for different values of imposed current $I$,
 smoothing length $\lambda$ (in units of the magnetic length), collision
 broadening $\gamma$, and temperature $T$. 
}
\end{minipage}
\end{figure}

So far we have considered situations in which, for sufficiently low $T$ and
small $\gamma$, we found well developed ISs. Figure~\ref{fig:RlHvsB} shows
that the corresponding $B$-fields ($\sim 7\,$T) are located in the
$\nu=2$ QH plateau near 
its high-$B$ edge. 
Before we consider other situations, we
recall a problem of the approach discussed so far \cite{Siddiki04:195335}.
As $B$ is lowered, the ISs move towards the sample edges and become narrower.
It is an artifact of our TFA that, similar to the
``perfect screening approximation'' of Refs.~\cite{Chklovskii92:4026} and
\cite{Chklovskii93:12605},  it yields ISs for any even-integer value of
$\nu(x)$, possibly with extremely small width,  provided the
temperature is low and the collision broadening is small enough. If, for
example, well developed ISs with local filling factor $\nu(x)=4$ exist (in our
case for $B\lesssim 3.6\,$T), the TFA predicts the existence of additional
narrow ISs with local filling $\nu(x)=2$ close to the edges. The local
equilibrium assumption requires that, in the limit $T\rightarrow 0$, $\gamma
\rightarrow 0$, a part of the imposed current flows dissipationless through
the ISs with  $\nu(x)=2$. While this has no effect on the vanishing of the
longitudinal resistance  $R_{\ell}$ in this limit, it leads for the
Hall resistance to a value larger than  $R_H=h/(4e^2)$, which is the value
expected in the absence of the ISs with $\nu(x)=2$. It is also an artifact of
the TFA that, in the limit $T\rightarrow 0$, $\gamma \rightarrow 0$,
the $\nu=2$ QH plateau extends down to $B$-fields at which ISs with local
$\nu(x)=4$ occur  (see black lines in Fig.~\ref{fig:RlHvsB}, $\lambda=0$). 
Of course, the accuracy of
our calculation becomes insufficient, if the width of ISs becomes comparable
with the separation of our mesh points.

 To overcome these artifacts of the TFA, we introduce the averaged
 conductivity tensor, 
Eq.~(\ref{aversigma})  \cite{Siddiki04:195335}. Then the ISs become
ineffective, if their width becomes less than $2\lambda$. In
Fig.~\ref{fig:RlHvsB} calculated values for Hall and longitudinal resistance
are shown for several values of temperature, collision broadening, and imposed
current, with $\lambda$ taken as a multiple of the magnetic length
$\ell_B$. With increasing $\lambda$ the ISs become ineffective at higher $B$
values, so the low-$B$ edge of the QH plateaus occurs at larger $B$.
Increasing temperature and collision broadening have a similar
effect. However, small collision 
broadening  leads to strongly reduced longitudinal
resistance in the tails of Landau levels, see Eq.~(\ref{sigma}).
 This leads to large effects on both sides of the QH plateaus,
whereas the other parameters have strong effects on the low-$B$ side of the QH
plateaus, but little effect on the high-$B$ side. With
increasing imposed current $I$ the relevant IS becomes wider and, if the other
parameters are fixed, the low-$B$ edge of the plateaus shifts to lower $B$
values.

\begin{figure}[t]
\noindent \begin{minipage}[b]{0.5\linewidth}
  \includegraphics[width=1.15\linewidth,angle=0]{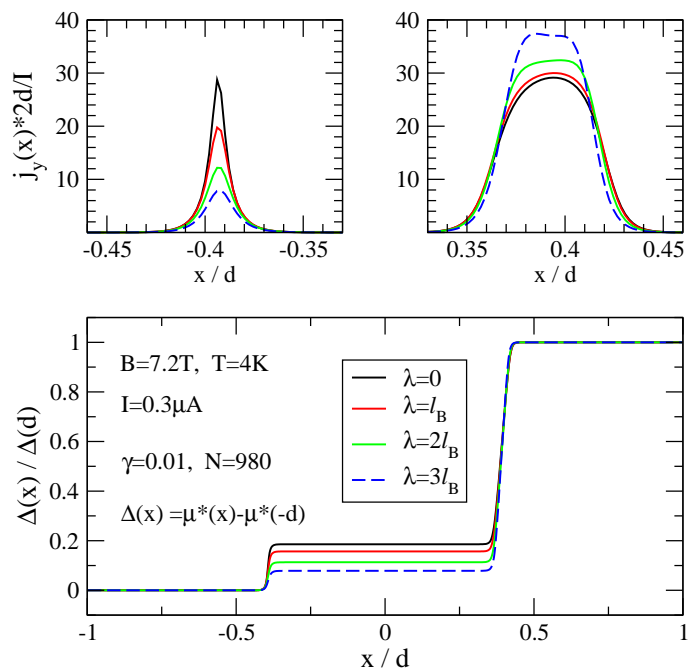}
\end{minipage} \hfill
 \begin{minipage}[b]{0.5\linewidth}
\caption{ \label{fig:z0-3b720}\small Effect of smoothing on current
  density and normalized Hall potential. $B=7.2\,$T.}
\end{minipage} 
\end{figure}
\begin{figure}[h]
\noindent \begin{minipage}[b]{0.5\linewidth}
  \includegraphics[width=1.15\linewidth,angle=0]{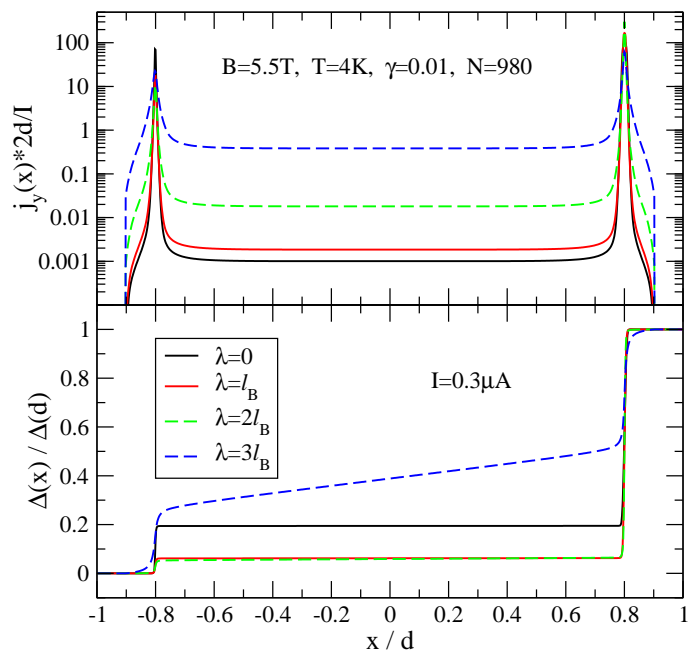}
\end{minipage} \hfill
 \begin{minipage}[b]{0.5\linewidth}
\caption{ \label{fig:z0-3b550} \small As  Fig.~\ref{fig:z0-3b720}, but
  for  $B=5.5\,$T.}\end{minipage}
\end{figure}

\begin{figure}[h]
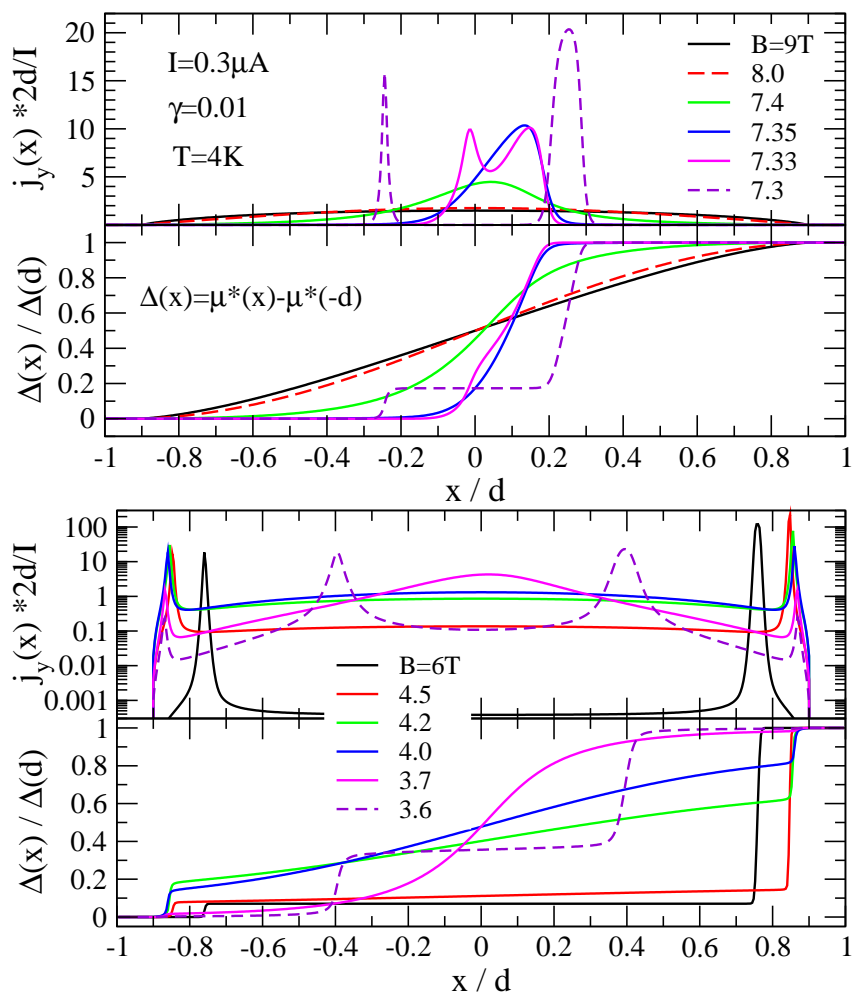

\noindent \begin{minipage}[b]{0.65\linewidth}
\includegraphics[width=1.1\linewidth,angle=0]{gerfig10.eps}\\
  \includegraphics[width=1.1\linewidth,angle=0]{gerfig11.eps}
\end{minipage} \hfill
\begin{minipage}[b]{0.35\linewidth}
\noindent
\caption{\noindent \label{fig:jyuh-9-36} \small Distribution of imposed current
  and normalized Hall potential for several $B$ values. $I=0.3\,\mu$A,
  $T=4\,$K, $\gamma=0.01$, $\lambda=\ell_B$.}\end{minipage}
\end{figure}

Depending on the width of the ISs, the smoothing of the conductivity can have
different effects. For relatively broad ISs, as in Fig.~\ref{fig:z0-3b720} for
$B=7.2\,$T, the smoothing just reduces the current through the narrow and
increases the current through the broader strip, and thus increases the
asymmetry. 
Figure~\ref{fig:z0-3b550} shows the corresponding results for $B=5.5\,$T,
where the ISs are considerably narrower. For small $\lambda$ the effect is
similar to that in  Fig.~\ref{fig:z0-3b720}, but for larger  $\lambda$ the
current density is reduced in both ISs and spreads out into the compressible
region, resulting in a finite Hall field  between the ISs and a breakdown of
the QHE.
An overview on the $B$-dependence of current distribution and variation of the
Hall potential is given in Fig.~\ref{fig:jyuh-9-36}.
At high magnetic fields, $B \gtrsim 8\,$T, only the lowest Landau level is
partly occupied, there exist no ISs, and the current is spread out over the
whole sample,
so that the Hall potential varies nearly linearly, as in the Drude limit. As
$B$ is lowered and the filling factor  in the center increases towards
$\nu(0)\approx 2$, the current concentrates more and more in the center
region, see the result for $B=7.4\,$T, until near $B=7.35\,$T an
incompressible strip occurs in the center, and the current is essentially
confined to that IS.
\begin{figure}[b]
\noindent \begin{minipage}[b]{0.55\linewidth}
\includegraphics[width=1.1\linewidth,angle=0]{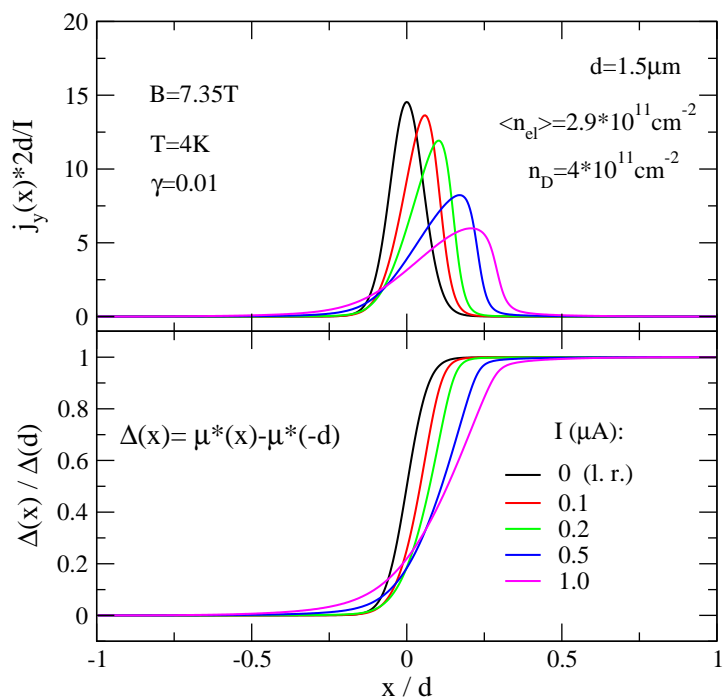}
\end {minipage} \hfill
 \begin{minipage}[b]{0.45\linewidth}
\caption{ \label{fig:jyuh-b735} \small Distribution of imposed current
  and normalized Hall potential for $B=7.35\,$T and several $I$ values.
  $T=4\,$K, $\gamma=0.01$, $\lambda=\ell_B$.
}\end {minipage}
\end{figure}
 With increasing current $I$, the peak of  $j_y(x)$
becomes increasingly asymmetric, with the maximum on the side where $I$ and the
intrinsic current density have the same direction, as shown in
Fig.~\ref{fig:jyuh-b735}.  As $B$ is lowered further,
the IS splits into two ISs, which rapidly separate spatially (see
Fig.~\ref{fig:jyuh-9-36} for $B\le 7.3\,$T).

\section{\label{sec:4} Discussion}

The current-induced asymmetry, which has already been mentioned in
Ref.~\cite{Guven03:115327}, is due to the existence of incompressible
strips, which develop at sufficiently low temperature and small collision
broadening.   
At higher temperature and larger collision broadening the
conductivity components are essentially proportional to the local density as
in the Drude limit. Then the Hall electric field becomes constant,
Eq.~(\ref{jE2}), and the Hall potential varies linearly over the sample. The
self-consistent confinement potential gets a corresponding linear
contribution, so that no major changes of the electron density result.
At high magnetic fields, $B \gtrsim 8\,$T, where only the lowest
Landau level is partly occupied, the situation is similar, even at
very low temperatures.

At low temperature and for small collision broadening, the local thermodynamic
density of states
is very large, where the electrochemical potential is close to a Landau level
(LL), and very small otherwise, and the sample approaches the ``perfect
screening'' situation \cite{Chklovskii92:4026,Chklovskii93:12605}.
In a typical equilibrium situation, as shown in Figs.~\ref{fig:v0dens} and
 \ref{fig:vnujy}, with  $\nu(0)\gtrsim 2$, in the
center region the LL $n=1$ is pinned to the constant $\mu^*$, while in the
outer regions near the edges, where $n_{\rm el}(x)$ drops to zero, the lowest
 LL, $n=0$, is pinned to $\mu^*$.
 These ``compressible'' regions are separated by ISs, in which 
 the eigenstates $|n,X\rangle$ carry the current $-e \langle
n,X|\hat{v}_y|n,X\rangle 
=(e/m\omega_c)dE_n(X)/dX$.  Since the potential energy varies across the ISs
by $\hbar \omega_c$, in the right IS flows the dissipationless current
$I_R=e\omega_c/\pi$ and in the left IS the opposite current $I_L=-I_R$. For
$B=7.2\,$T, the Hall voltage across the right IS is $U_R=\hbar
\omega_c/e=12.46\,$mV corresponding to a current
$I_R=U_R/R_H(2)=0.965\,\mu$A. If now an additional stationary 
 current $I^{ad}$ is imposed, the general rules of non-equilibrium
 thermodynamics require that the system assumes the stationary state with
 minimal entropy  production, i.e. the additional current has to flow also
 dissipationless through the incompressible strips. For $I^{ad}>0$ the current
 part 
 $I_R^{ad}>0$ will enhance the current through the right IS to $I_R+I_R^{ad}$ and
 the voltage drop across it to $U_R+R_H(2)I_R^{ad}$. The part
 $I_L^{ad}=I^{ad}-I_R^{ad}>0$ 
reduces the intrinsic current through  and the voltage drop across the
left IS. Electrostatic arguments  \cite{Chklovskii92:4026} show that the width
$w$ of an IS increases with the potential drop $\Delta V$ across it. 
Chklovskii {\em et al.} estimate $w^2=a\cdot \Delta V$, where the constant $a$
is inversely proportional to the slope of the (zero-$B$) density profile in
the center of 
the IS  [see Eq.~(20) of Ref.~\cite{Chklovskii92:4026}]. This predicts,
that the IS, in which intrinsic and imposed currents have the same direction,
becomes wider, whereas the width of the other IS, where the imposed current 
reduces the intrinsic one, is reduced. This prediction is confirmed by the
self-consistent calculations. These show further, that the part $I_L^{ad}$ of the
imposed current flowing in the left IS, becomes
equal to the other part $I_R^{ad}$ in the linear response limit of small $I^{ad}$,
whereas it becomes much smaller, $I_L^{ad} \ll I_R^{ad}$, for large $I^{ad}$.
At higher temperature and for finite collision broadening, screening is not
perfect and the ``perfect screening'' results hold only approximately.
These non-linear transport effects are most important near the low-$B$ edges
of the QH plateaus.

Also the  smoothing of the conductivity tensor is most effective near
these edges.
This smoothing has a strong effect on narrow ISs, where it removes the
zeroes of $\sigma_{\ell}(x)$, but only a weak effect on wide ISs, where it
mainly modifies the behavior near the edges. Thus it  enhances
the current-induced asymmetry, as is shown in Fig.~\ref{fig:z0-3b720}.
 As discussed in
Ref.~\cite{Siddiki04:195335} for the linear-response limit of weak
imposed current, replacing of $\hat{\sigma} (x)$ by its average
$\hat{\bar{\sigma}}(x)$ over an interval of length $2\lambda$ reduces the width
of the QH plateaus, as shown in Fig.~\ref{fig:RlHvsB}.
Taking the averages with sufficiently large $\lambda$, makes the ISs with
$\nu(x)=2$ ineffective already at higher $B$-fields, for which no ISs with
$\nu(x)=4$ exist in the sample. Then the imposed current is spread out to
the compressible region and suffers dissipation. Consequently, the QH plateau
with $\nu=2$ shrinks at its 
low-$B$ side, where  $R_H$ decreases and $R_{\ell}$ becomes finite. This is
shown in Fig.~\ref{fig:RlHvsB} for $\lambda= \ell_B$ and $\lambda=2 \ell_B$.
We see that for finite imposed current the effects of  $\lambda$, of the
collision broadening described by $\gamma$, and of temperature are
qualitatively similar as in the linear response regime $I\rightarrow 0$. In
addition Fig.~\ref{fig:RlHvsB} shows that the current-induced asymmetry of the
ISs, resulting in a broadening of one IS,  tends to broaden the QH
plateaus at their low-$B$ edge and, consequently, to lower $R_{\ell}$.

 The high-$B$ edge of the $\nu=2$ QH plateau seems rather
insensitive to changes of all these parameters except $\gamma$: if the
collision broadening is reduced, $R_{\ell}$ becomes
smaller. Nevertheless, as  Fig.~\ref{fig:jyuh-b735} shows, the imposed
current induces an asymmetry and a broadening of the region, in which the
Hall potential varies, even if the filling factor in the center is only
slightly below 2, and the current is confined to the center region.
This asymmetry disappears, however, rapidly as $B$ increases above
 the high-$B$ edge of the plateau, as can be seen from the results
for $B=7.4\,$T and $B=3.7\,$T in Fig.~\ref{fig:jyuh-9-36}.

Our results are in reasonable agreement with the experiments
\cite{Panos11:thesis,Panos13:unpub}, where two cases could be
distinguished. Near the  
low-$B$ edges of the QH plateaus an ``edge-dominated behavior'' was found, in
which most of the Hall potential dropped in an asymmetric way over relatively
narrow stripes near the edges of the Hall bar. Near the high-$B$ edges of the
QH plateaus
 a ``bulk-dominated behavior'' was observed,
with the major change of the Hall potential near the center of the Hall
bar. The exact position of the potential drop changed along the bar, probably
due to inhomogeneities. Such  inhomogeneities make our assumption of
translation invariance a crude approximation to reality, and
may be the reason, why in the experiments the bulk-dominated
situation is observed in a wider interval of $B$-values than in our results.

We find in the edge-dominated situations that the asymmetry of the Hall
voltages across the ISs is connected with a difference in the widths of the
ISs, notably with a remarkable broadening of the IS, which caries the larger
current and leads to the larger step in the Hall potential. Whereas the
asymmetry of the potential steps is clearly seen in the experiments, the
current-dependence of the width of the corresponding strips is not clearly
observed (see Fig.~\ref{fig:edgedom}), possibly because of the restricted
spatial resolution ($\sim 100\,$nm). The weak asymmetries seen in
Fig.~\ref{fig:jyuh-b735} do not contradict to the experimental results for
weak imposed currents, but can also not clearly be confirmed by the experiments
 \cite{Panos11:thesis,Panos13:unpub}. At higher currents ($I\approx 3.9\,\mu$A)
 the 
 experimental curves change  suddenly their shape \cite{Panos11:thesis},
 probably due to the breakdown of the QHE.

The breakdown of the QHE with inreasing current can also be seen from
Fig.~\ref{fig:edgedom}, which shows that with increasing $I$ the slope of the
Hall potential in the compressible region between the ISs increases. Thus, a
part of the current flows through the dissipative bulk, and the resistance is
no longer quantized. Our results, on the other hand, say that with
increasing $I$ one of the ISs becomes wider, carries more current
dissipationless, and leads to an extension of the QH plateau at its low-$B$
edge. This points to a serious shortcoming of our approach: it does not
consider breakdown effects, which may become relevant at high imposed
currents. One of these effects is Joule heating, which will become important
at these low temperatures, if the current flows, at least partly, through
compressible, i.e. dissipative regions. Another one is the often discussed
quasi-elastic inter-Landau-level-scattering (QUILLS)
\cite{Eaves86:346,Guven02:155316}, which must become important if, in strong
in-plane electric fields, isoenergetic states of adjacent Landau bands tend
to overlap spatially. Exactly this happens at the wide IS with the large
potential step $\Delta V$ and the width $w\propto \sqrt{\Delta
  V}$. There the Landau energies have a slope $dE_n(X)/dX \approx \Delta V/w$,
so that the spacing of the states $|0,X_0\rangle$ and $|1,X_1\rangle$
with equal energies $E_0(X_0)=E_1(X_1)$ is $|X_0-X_1|=\hbar \omega_c w/\Delta
V \propto 1/ \sqrt{\Delta V}$, and becomes small as  $\Delta V$ increases with
increasing current $I$.

Apart from these well understood shortcomings of the existing calculations,
the self-consistent screening theory provides a good understanding of the
experimental results. The observed difference between ``edge-dominated
behavior'' near the low-$B$ edges of the QH plateaus and the ``bulk-dominated
behavior'' near the high-$B$ edges becomes clear from Figs.~\ref{fig:RlHvsB}
and \ref{fig:jyuh-9-36}. The edge-dominated behavior results from the two
separated 
incompressible strips, which become narrower and less effective with
decreasing $B$. But near the  low-$B$ edges of the QH plateaus, the ISs
dominate the transport and, at fixed finite temperature, the longitudinal
resistance decreases exponentially with increasing $B$ throughout the whole
plateau region (see Fig..~\ref{fig:RlHvsB}).
 The bulk-dominated behavior, on the other hand, is determined
by the $B$-value at which the population of a new Landau level starts. At
slightly lower $B$-values an incompressible region exists in the center. At
slightly larger $B$-values only the lower Landau level is nearly filled in the
center of the sample, so that the resistivity there is very small. In both
cases the center region determines the transport, which therefore is
``bulk-dominated''. 

A basic assumption of our explanation of the QHE in narrow Hall bars is, of
course, that the lateral confinement 
of the 2DES introduces a position-dependence of the electron density, which
decreases from a maximum  in the center to zero near the edges of the
sample. 

\ack
We are grateful to Klaus von Klitzing for his continuing interest in
and his support of this work.

\section*{References}


\end{document}